\documentclass{aa}  

\usepackage{graphicx} 
\usepackage{txfonts} 
\usepackage{tabularx} 

\usepackage{xcolor}
\begin{document}

   \title{Calibrating echelle spectrographs with Fabry-P\'{e}rot etalons}

   \author{F. F. Bauer, 
          M. Zechmeister 
          \and A. Reiners 
          } 

   \institute{Institut f\"ur Astrophysik (IAG), Georg-August-Universit\"at G\"ottingen, 
              Friedrich-Hund-Platz 1, D-37077 G\"ottingen\\ 
              \email{fbauer@astro.physik.uni-goettingen.de} 
             } 

   \date{submitted} 

 
  \abstract 
   {Over the past decades hollow-cathode lamps have been calibration standards for spectroscopic measurements. Advancing to cm/s radial velocity precisions with the next generation of instruments requires more suitable calibration sources with more lines and less dynamic range problems. Fabry-P\'{e}rot interferometers provide a regular and dense grid of lines and homogeneous amplitudes making them good candidates for next generation calibrators.} 
   {We investigate the usefulness of Fabry-P\'{e}rot etalons in wavelength calibration, present an algorithm to incorporate the etalon spectrum in the wavelength solution and examine potential problems.} 
   {The quasi periodic pattern of Fabry-P\'{e}rot lines is used along with a hollow-cathode lamp to anchor the numerous spectral features on an absolute scale. We test our method with the HARPS spectrograph and compare our wavelength solution to the one derived from a laser frequency comb.} 
   {The combined hollow-cathode lamp/etalon calibration overcomes large  distortion ($50$ m/s) in the wavelength solution of the HARPS data reduction software. Direct comparison to the laser frequency comb bears differences of only maximum $10$ m/s.} 
   {Combining hollow-cathode lamps with Fabry-P\'{e}rot Interferometers can lead to substantial improvements in the wavelength calibration of echelle spectrographs. Etalons can provide economical alternatives to the laser frequency comb, especially for smaller projects.} 


   \maketitle 
%

\section{Introduction} 

Accurate wavelength calibration is a cornerstone for any measurement with high resolution spectrographs. It translates the pixel position of the detector to absolute wavelengths and thereby defines the physical scale. As many astronomical research fields benefit from high precision radial velocity (RV) measurements, the next generation of spectrographs aims for cm/s precisions \citep{2007MmSAI..78..712D}. Today, most echelle spectrographs use hollow-cathode lamps (HCLs), e.g.,  Thorium-Argon or Thorium-Neon, to perform wavelength calibration but these lamps are insufficient for the needs of future high precision instruments. 

Hollow-cathode lamps establish collisions between the carrier gas (Argon or Neon) and the Thorium atoms which thereby get exited \citep{2007ASPC..364..461K}. When the Thorium relaxes into a lower energy state, a photon carries away the energy difference and contributes to an emission line. The laboratory wavelengths of Thorium lines can be measured or calculated as done by \citet{1983ats..book.....P}, \citet{2007A&A...468.1115L}, \citet{2008ApJS..178..374K} and \citet{2014ApJS..211....4R}. The drawback of HCLs is the limited number of reference lines, the large dynamic range spanned by the lines, and the irregular line distribution. Blending of lines is an additional problem: in total only 4000 – 8000 spectral lines remain usable in the visual \citep{2007A&A...468.1115L}. In addition to the large intensity differences between Thorium lines, the carrier gas often emits strong lines itself. Exposure times have to be well chosen in order to have enough cathode (e.g. Thorium) lines for calibration while keeping the number of saturated carrier gas (e.g. Argon) lines low. Spectral regions without lines must be interpolated by the wavelength solution resulting in uncertain wavelengths for these pixels. 

To overcome the problems of HCLs new calibrators are currently under development. Among them are the laser frequency comb (LFC, e.g. \citealp{2007MNRAS.380..839M,2010MNRAS.405L..16W,2012Natur.485..611W}) and the Fabry-P\'{e}rot interferometer (FPI, \citealp{2010SPIE.7735E..4XW}; \citealp{2012SPIE.8446E..94S}; \citealp{2014A&A...569A..77R}; \citealp{2014arXiv1404.0004S}). LFCs are rather expensive and often not affordable for smaller projects. Here we focus on less cost intensive FPIs and their use as a calibration source. 

A tunable FPI device was already used as dispersing element to develop a high resolution spectrograph for planet search located at Kitt Peak \citep{1988ASPC....3..237M, 1994Ap&SS.212..271M}. The wavelength solution of the spectrograph relied totally on geometrical calibration and accurate tuning of the FPI demonstrating precisions of $6$ m/s. \citet{1982SPIE..331..315C} used the FPI to superimpose absorption lines of known wavelength in the stellar spectrum. Both used the FPI in transmission resulting in very low instrument efficiency leading to the development of new techniques such as iodine cells in transmission \citep{1996PASP..108..500B} or ultra stable spectrographs \citep{2003Msngr.114...20M}. Currently, FPIs are again considered in astronomical spectrographs for nightly drift checks, but not as wavelength calibrators \citep{2010SPIE.7735E..4XW}. 

The advantage of the FPI against HCLs is its ability to produce a dense grid of lines with almost uniform intensity over the entire spectral range. They can be manufactured to match the specifications of most spectrographs as the free spectral range is a function of the cavity width. This tackles the problem of empty spectral regions and saturated lines thereby increasing the robustness of the wavelength solution. 

The drawback of FPIs is their weakly constrained absolute wavelength. Line positions are determined by the cavity width of the etalon and the interference order. Both are usually not known to the accuracy needed to provide absolute calibration. Furthermore, the group-velocity dispersion limits the predictability of individual peak positions, and the cavity width is very sensitive to temperature and mechanical stress. Hence FPIs are temperature and pressure stabilized to minimize drifts (\citealp{2010SPIE.7735E..4XW}; \citealp{2012SPIE.8446E..94S}). Such a stabilized FPI is located at the HARPS spectrograph and nightly drifts of $10$\,cm/s were reported (\citealp{2010SPIE.7735E..4XW}).

First, we propose a method to characterize the FPI comb pattern with HCLs, using a high resolution spectrograph in Sect~2. We explain our wavelength calibration concept and model in Sect.~3. In Sect.~4 we combine information from an HCL and a FPI to improve the accuracy and stability of the HARPS wavelength solution as compared to the use of an HCL alone. Furthermore, we investigate the impact of a distorted wavelength solution on RV measurements in Sect.~5

\section{Calibrating the Fabry-P\'{e}rot interferometer} 

\subsection{The ideal FPI} 

\begin{figure} 
\includegraphics [width=\linewidth]{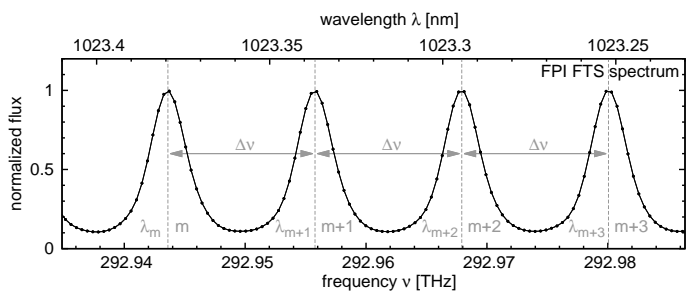} 
\caption[Schematic FPI]{Spectrum of the CARMENES NIR FPI taken with the Fourier Transform Spectrometer in G\"ottingen.} 
\label{fig:schematic_FPI} 
\end{figure}  

The ideal FPI consists of two partly reflective plane parallel surfaces \citep{1999prop.book.....B}. When light enters the device it  gets reflected between the two surfaces and interference takes place between light rays depending on the phase difference. With a white light input source, a spectrograph will observe a pattern of transmission  peaks spaced equidistantly in frequency $\nu$ (Fig.~\ref{fig:schematic_FPI}). The transmission maxima fulfill the interference condition given by the effective FPI cavity width $d$:  
   \begin{equation} 
      m \lambda_m = 2d 
      \label{eq:FPI} 
   \end{equation} 
   where $m$ is an integer number called interference order and $\lambda_m$ are the transmission maxima wavelengths. The effective cavity width is $d=D\,n\,\cos \theta$ with the distance between the two mirrors $D$, the refraction index $n$ and the incident ray angle $\theta$. 
 
 Using the FPI as a calibration source requires knowledge of the exact wavelengths for all  interference maxima observed by the spectrograph. As the wavelength $\lambda_m$ of each FPI interference order  depends only on the effective cavity width, $d$, this parameter must be known with the same accuracy as the wavelength requirement ($\delta d / d = 3 \cdot 10^{-9}$ for $1$ m/s). For the FPI this means that the spacing $d$ must be known with an accuracy of about $0.15~\AA$ \citep{2014A&A...569A..77R}. Typically the mirror distance $D$ is only known to about $1~\mu$m which is five orders of magnitudes above our requirements. In addition we do not exactly know the incident ray angle $\theta$. Hence, we need to find a method to calibrate the effective FPI cavity width $d$. 
  We use the interference condition of  the FPI, Eq.~(\ref{eq:FPI}), to perform the cavity width measurement. This can be done with any high resolution spectrograph and its internal calibration sources such as HCLs \citep{2010SPIE.7735E..4XW}. First, we derive the wavelength solution of the spectrograph using only HCL lines. Then we derive the line positions of the FPI on the detector and use the HCL wavelength solution to assign the corresponding wavelength to the interference peaks. 
 
Now we need to identify the interference order $m$ for all FPI peaks. From the observed spectra we can only obtain a relative numbering $k = m - m_1$ by simply counting the lines but the absolute interference order of the first (reddest) peak $m_1$ is not known. To find $m_1$ we use the fact that the ideal FPI cavity width is a constant number. We start with a guess for $m_1$ and use the resulting interference orders $m = k + m_1$ and the wavelengths derived for each line in Eq.~(\ref{eq:FPI}) to obtain the effective cavity width $d$ for each FPI line separately. If we plot the effective cavity width $d$ as a function of relative peak number $k$ using the correct $m_1$ we expect the data points to be distributed around a constant value. If our guess for $m_1$ is wrong the data points will, however, produce a positive slope for $m_1$ too small or a negative slope for $m_1$ too high. Once the right value for $m_1$ is found the large number of individual effective cavity width measurements (typically in the order of $10^4$ for high resolution echelle spectrographs) can be averaged to obtain a good estimate of $d$. 
 
We now only need this one parameter (the effective cavity width $d$ of the FPI) to globally predict the wavelength of all interference peaks observed. The combined accuracy of all HCL lines anchors the FPI on an absolute wavelength scale. Paired with the high precision of the FPI local imperfections of the wavelength solution can now be resolved and corrected. The numerous FPI features bridge regions lacking any spectral features of HCLs and we can now check if the wavelength solution model is sufficient or if more detailed models are required. 

\subsection{The real FPI} 

In practice, applying the concept of calibrating the effective cavity width using standard calibrators proves to be more difficult for real FPIs. The cavity width is not constant throughout the wavelength range covered by high resolution echelle spectrographs. FPIs are usually soft coated allowing photons of different energy to penetrate to different depth  of the dielectric surface. Hence, photons of different wavelength see different cavity widths. As an example the penetration depth variation over the wavelength range of HARPS easily reaches  hundred nm \citep{2010SPIE.7735E..4XW}. If we do not account for this the computed FPI wavelengths are incorrect by several km/s. For a full characterization of the FPI spectrum, the task is therefore not only to find the global mean effective cavity width $\bar{d}$ of the FPI but also to determine the wavelength dependent function  $d(\lambda)$. 

We assume that the penetration depth is a smooth function of wavelength and therefore also a smooth function of relative interference order $k$. Because the measured wavelength is uncertain but the relative peak numbering is noise-free, we model $d$ as a function of $k$ rather than $d$ as a function of $\lambda$. We choose uniform cubic B-splines (\citealp{de2001practical}; \citealp{dierckx1995curve}) to fit the function $d(k)$. 

The assumption that the penetration depth variation is smooth is the cornerstone of our method. Hence we decided to test this with  the CARMENES \citep{2011IAUS..276..545Q} near infrared FPI (finesse $\sim 8$, \citealp{2012SPIE.8446E..94S}). We obtained 78 spectra with the Bruker IFS 125/HR Fourier transform spectrometer (FTS) with a resolution of $R=500\,000$ in our optics laboratory. The median signal-to-noise ratio of the coadded spectrum is 350 (Fig.~\ref{fig:schematic_FPI}). As starting guess for all FPI peak wavelengths we used the internal wavelength calibration of the FTS. It is provided by a He-laser measuring the path length within the spectrometer. With the high resolution of the FTS the FPI peaks are well resolved in the spectra and we measured individual line positions by fitting a Lorentz function. The FPI itself was not operated under vacuum conditions, thus water lines contaminated the spectrum. We excluded contaminated peaks from our cavity width measurement. We determined the mean effective cavity width $\bar{d}=12.338\,350 \pm 0.000\,040$~mm which is consistent with what we expect from the mirror distance given by the manufacturer $D=12.334$~mm, the refractive index of air $n \sim 1.003$, and an incident ray angle of $\theta=0$~deg. 

\begin{figure} 
\includegraphics [width=\linewidth]{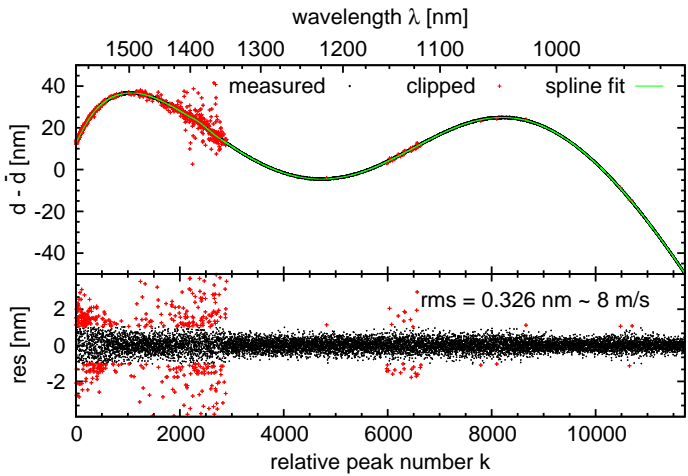} 
\caption[D_CARMENES]{Upper panel: Effective cavity width measurement $d(k)$ for the CARMENES NIR FPI. Lower panel: Residuals of spline fit. Data (black dots), clipped lines due to atmosphere absorption features (red crosses), spline fit to the data (green line).} 
\label{fig:D_CARMENES} 
\end{figure}

The penetration depth variation of the FPI as a function of the relative peak numbering and the measured FTS wavelength is shown in Fig.~\ref{fig:D_CARMENES}. Fitting a B-spline with $50$ nodes results in an RMS scatter around the model of $0.326$ nm translating to 8 m/s. We could not reveal any discontinuities in $d(k)$ within our measurement precision. Hence our assumption that the penetration depth variation is smooth is valid. The RMS scatter of 8 m/s is about what we expect from photon noise. 

Once we have determined the model parameters for the effective cavity width $d$ we need to find the absolute interference number of first observed FPI peak $m_1$ to compute the wavelength for all interference orders. This is no longer trivial if the effective cavity width $d$ is not a constant. 

\citet{2014A&A...569A..77R} suggested  to use the mean free spectral range $\langle \Delta \nu \rangle$ to calculate the mean effective cavity width and then use Eq.~(\ref{eq:FPI}) to calculate the absolute order  numbering.  \citet{2010SPIE.7735E..4XW} determined the absolute order numbering assuming the effective cavity width $d$ to be exactly the absolute cavity width $D$ specified by the manufacturer and rounding $m$ to the closest integer number they  obtain from equation Eq.~(\ref{eq:FPI}) and the HARPS wavelength solution. Their approach resulted in the penetration depth curve with the minimum variation. 

\begin{figure} 
\includegraphics [width=\linewidth]{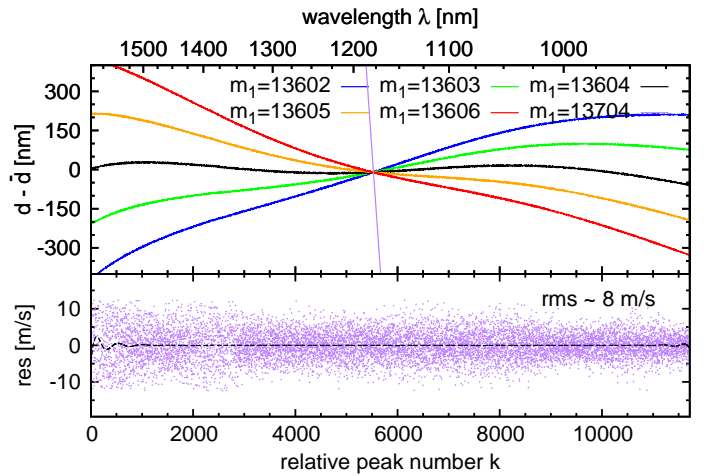} 
\caption[D_CARMENES]{Upper panel: Effective cavity width $d(k)$ obtained for different absolute interference order numbers $m_1$. The reference curve ($m_1=13604$, black) is shown in Fig.~\ref{fig:D_CARMENES}. Lower panel: Difference in modeled FPI peak wavelength $\lambda_m$ using $m_1=13704$ with the reference curve (dashed line). Residuals of the spline fit to the cavity width $d(k)$ obtained with $m_1=13704$ (purple dots).} 
\label{fig:different_m} 
\end{figure}

 Other choices of $m_1$ introduce steeper slopes in the effective cavity width $d$. We show how the choice of $m_1$ influences the overall slope of the function  $d(k)$ for the CARMNENES NIR FPI  in Fig.~\ref{fig:different_m}. Because we do not know enough about the material properties of the coating, the penetration depth variation with wavelength and the alignment of the FPI, we do not know which $m_1$ and which $d$ curve are the true ones. We chose $m_1=13604$ as reference value because it results in the $d(k)$ curve with the minimal penetration depth variation. We note that, this choice is not necessarily the true value of $m_1$ 
 
 We realized, however, it is not necessary to know $m_1$ exactly to calibrate the FPI wavelengths.  Due to the degeneracy in Eq.~(\ref{eq:FPI}) between the absolute peak number $m$ and the effective cavity width $d(k)$, identical peak wavelengths occur for different combinations of $m$ and $d(k)$. Hence, our method to derive absolute wavelengths for individual FPI peaks should be robust against choosing the wrong value for $m_1$. 
 
 To test this we computed $d(k)$ using two different values for $m_1$ differing by $100$ interference orders, $m_1=13604$ and $m_1=13704$. For both cases we obtain the FPI wavelengths using Eq.~(\ref{eq:FPI}). We calculate the difference of the calibrated FPI wavelengths line by line and show the result as dashed line in the lower panel of Fig.~\ref{fig:different_m}. The maximum difference between the calibrated peak wavelengths is $10$~cm/s in the region between $k=2000$ and $k=10000$. At the edges the difference increases up to $5$~m/s due to the decreasing number of data points to constrain the spline fit. Despite the large difference in $d(k)$ (several hundred nm, see upper panel in Fig.~\ref{fig:different_m}) the difference between the modeled wavelengths is small (lower panel in Fig.~\ref{fig:different_m}).

%

\section{Wavelength solution model}

In this section, we explain our strategy to determine the wavelength solution of a high-resolution echelle spectrograph. First, we introduce our concept to carry out the mapping between wavelength and pixels in the case of a monolithic detector with uniform pixels. We motivate our model, parametrization and regression type. Second, we advance to real detectors and show how we account for uneven pixel sizes in the wavelength solution. 

\subsection{Mapping wavelengths to detector coordinates}

The general idea of wavelength calibration is to feed the spectrograph with a reference spectrum of known wavelengths $\lambda_l$, for example the known atomic lines of a hollow-cathode lamp, and observe the response at the detector. The positions of the line centers $x_l$ in the extracted spectrum are measured by fitting a model of the instrumental profile (e.g. Gaussian functions) to each individual spectral line $l$.
For the set of wavelength-pixels pairs $(\lambda_l,x_l)$, a relation is established via a fit. Finally, this relation is used to predict the wavelength at the center of every detector pixel.

The wavelengths of standard lines are given by accurate and precise line lists with uncertainties in the range of $10^{-6}-10^{-7}$ ($\sim\,10$\,m/s) \citep{2014ApJS..211....4R}. On the other hand, the positions of weak lines on the detector exhibit typical measurement errors of $0.1$ pixels or about $100$\,m/s.
Hence, in wavelength calibration the standard line wavelength, $\lambda_l$, represents the independent variable (or the cause) and the position on the detector, $x_l$, is the dependent variable (or the spectrograph response).

A common and simple procedure to do wavelength calibration is to fit a continuous model for $\lambda(x)$. From the function $\lambda(x)$ one can evaluate directly the wavelength for each pixel. In this approach the role between dependent and independent variable is reversed \citep{1990ApJ...364..104I} and we call this reverse regression.
Reverse regression causes problems when weighting the data.

For this reason one could use direct regression and fit the detector position of individual lines as a function of wavelength, $x(\lambda)$.
To predict the wavelength for each pixel in this approach the model $x(\lambda)$ must be inverted $\lambda(x) = x(\lambda)^{-1}$.

The wavelength solution is usually modeled with polynomials which, in case of reverse regression is $o\lambda = \mathrm{poly}(x)$ (e.g. \citet{1996A&AS..119..373B}), can be interpreted as a Taylor expansion of the grating equation
   \begin{equation}
      o \lambda = n \sigma~[\cos \gamma_1~\sin \alpha_1+\cos \gamma_2~\sin \alpha_2]
      \label{eq:grating}
   \end{equation}
in terms of the diffraction angle $\alpha_2$ which is proportional to the position in the focal plane, $x\propto\alpha_2$.
In Eq.~(\ref{eq:grating}), $o$ is the diffraction order, $n$ is the refractive index, $\sigma$ is the grating constant, $\alpha_1$ and $\alpha_2$ are the angles of the incident and diffracted ray with respect to the grating normal, and $\gamma_1$ and $\gamma_2$ are the off-plane angles before and after the
grating, resp. \citep{2000asop.conf.....S}.

In direct regression with $x(\lambda) = \mathrm{poly}(o\lambda)$ we basically approximate a re-arranged version of the grating equation
   \begin{equation}
  \alpha_2 = \arcsin \Big( \frac{1}{\cos \gamma_2} \Big[\frac{o \lambda}{n \sigma} - \cos \gamma_1~\sin \alpha_1 \Big] \Big) \ .
      \label{eq:grating2}
   \end{equation}
We note that polynomials of higher degree (typically +1) are needed with direct regression for a wavelength solution with similar quality compared to reverse regression.
This is because the sine function in Eq.~(\ref{eq:grating}) is more amenable to a Taylor expansion than the arcsine function in Eq.~(\ref{eq:grating2}).

A way of keeping both advantages (lower number of parameters and direct regression) would be to use $x = \mathrm{poly}(o\lambda)^{-1}$ as forward model which we might call inverse regression. This requires non-linear least square fitting and thus higher computational effort.

In what follows we chose direct regression. This allows us to correctly weight all data points and to implement unequal pixel sizes in our model (next section).
Moreover, instead of simple polynomials for each order, we use 2D polynomials to couple the individual orders. This decreases the number of parameter and increases the robustness of the fit.  We also use $o\lambda$ instead of $\lambda$ as variable, i.e. $x(\lambda,o) = \mathrm{poly}(o\lambda, o)$. The re-parametrization is motivated by Eq.~(\ref{eq:grating2}). This approach is a direct regression version of the algorithm implemented in the IDL REDUCE package of \citep{2002A&A...385.1095P}.

\subsection{Unequal pixel sizes}

Standard wavelength calibration procedures usually assume that detector 
pixels are equally spaced and of the same size. Irregularities in pixel 
size of CCD detectors are on the order of $10^{-2}$ 
\citep{2010MNRAS.405L..16W}, which is becoming a significant limitation 
for high-precision spectroscopic measurements. For example, 
\citet{2010MNRAS.405L..16W} showed that inhomogeneous pixel sizes lead to 
discontinuities in the wavelength solution of HARPS on the order of 
several $10$\,m/s.

In order to account for irregular detector pixels, we include the pixel size 
in our model and we distinguish now more strictly between \emph{detector} 
coordinates, $X$ (pixel), and \emph{focal plane} coordinates, $x$. We 
denote the size of pixel $i$ by $a_{\mathrm{pix},i}$ and we assume that there is no gap 
space between individual pixels. Thus, the transformation from pixel coordinates to focal plane 
coordinate is
   \begin{eqnarray}
      x(X,\boldsymbol{a_\mathrm{pix}}) & = & \sum_{i=1}^{[X]} a_{\mathrm{pix},i} + a_{\mathrm{pix},[X]} \cdot \left( 
X-[X]-\frac{1}{2} \right)
\label{eq:coord_transformation}
   \end{eqnarray}
where $[X]$ is the value of $X$ rounded to the nearest integer (the 
pixel number). In Eq.~(\ref{eq:coord_transformation}) we integrate over the width of all 
preceding pixels (including the current one) and linearly interpolate (backwards) for sub-pixel position.
The zero point of the coordinate system ($x = 0$) coincides with the left 
border of the first pixel ($X = 0.5$). If all pixels have the same size 
($a_{\mathrm{pix},i} = 1$) the transformation is simply $x = X - 0.5$. The 
transformation can 
also be extended to include virtual pixels for the application to mosaic 
detectors with gap sizes corresponding to several pixels ($a_{\mathrm{pix},i}\sim100$\,pixels) 
between the detectors as, e.g., in CARMENES \citep{2011IAUS..276..545Q} or 
CRIRES+ \citep{2014SPIE.9147E..19F}.

Ideally, the sizes of all pixels are known prior to wavelength 
calibration, but in practice the individual pixel sizes are not known 
precisely enough. For future detectors, it might be interesting to 
determine pixel sizes with a microscope prior to integration in the 
instrument.

Determining the sizes of all pixels is beyond the scope of ordinary 
wavelength calibration because the number of free parameters exceeds the 
number of comb lines which are separated by a few pixels. But for a few, 
significantly deviating pixels it is indeed possible to simultaneously 
determine pixel sizes and the wavelength solution, if a spectral calibration source with dense line combs (FPI or the LFC) is available.

For example the HARPS detector exhibits pixel irregularities every 512 pixels \citep{2010MNRAS.405L..16W}, and we therefore choose the following pixel size model
\begin{eqnarray} 
\label{eq:pixsizes} 
             a_{\mathrm{pix},i} & = & \begin{cases}  
     w_i , & \text {if modulo}(i,~b \cdot 512 - 3) = 0 \\ 
     1 , & \text{otherwise} 
     \end{cases} 
\end{eqnarray}
where every $512$th pixel is allowed to depart from unity, while all others are assumed to be homogeneous and are fixed to unity. The first pixel size irregularity, $b=1$, is located in pixel $509$ motivating an offset of 3 pixels in Eq.~(\ref{eq:pixsizes}). One can further limit the number of free parameters when using column sizes instead of individual pixels as we do in Sect.~4 for HARPS.

To find the pixel sizes $w_i$, we perform a non-linear least-square fit by minimizing $\chi^2 = \sum_{l} \frac{1}{\sigma_l^2}[X_l - X(x(\lambda_l,o_l,\boldsymbol{a_{\mathrm{poly}}}),\boldsymbol{a_{\mathrm{pix}}})]^2$.
To evaluate the $\chi^2$ with the observed data $X_l$ we must invert Eq.~(\ref{eq:coord_transformation}). The back-transformation of the focal plane coordinates $x$ into the detector coordinates $X$ is 
  \begin{equation}
      X(x,\boldsymbol{a_{\mathrm{pix}}}) = \frac{x -x_j }{a_{\mathrm{pix},j}} + j + \frac{1}{2} 
      \label{eq:transformation_back} 
   \end{equation} 
 where $j$ is the pixel number into which $x$ falls which can be found via the condition 
 $x_j \leq x < x_{j+1}$ where $x_j = x(j-0.5,\boldsymbol{a_{\mathrm{pix}}}) = \sum _{i=1}^{j-1} a_{\mathrm{pix},i}$.

The best fitting model and coordinate transformation is then inverted numerically to obtain the wavelength for the center of each pixel. We found that a simple  bisection method is appropriate \footnote[2]{Methods involving first derivatives are not recommended by us because of convergence problems with discontinuous functions as it is the case here.}. 

To accelerate the non-linear least-square fit we separate it into a linear and a non-linear least-square fit. For a trial set of the non-linear parameters $\boldsymbol{a_\mathrm{pix}}$, the position of all calibration lines can be transformed to focal plane coordinates $x_l(X_l)$ with Eq.~(\ref{eq:coord_transformation}). Then a smooth 2D polynomial is fitted with a linear least-square fit. The smooth model is transformed back to detector coordinates with Eq.~(\ref{eq:transformation_back}) and the $\chi^2$ with the observed data $X_l$ is computed. Testing different trial sets (e.g. with a downhill simplex; \citet{press2007numerical}) the best fitting pixel sizes $\boldsymbol{a_\mathrm{pix}}$ are found. 

\section{Calibrating HARPS using its FPI} 

 \subsection{Detector characterization and wavelength solution} 

To show the potential of Fabry-P\'{e}rot interferometers in wavelength calibration procedures for echelle spectrographs, we calibrate the HARPS spectrograph located at the ESO 3.6 m telescope in La Silla Chile \citep{2003Msngr.114...20M} with its etalon and our wavelength solution method. 
HARPS has a resolution of $115\,000$ (one resolution element is sampled with about 3.2 pixels), covers the wavelength range from 380 to 690 nm and offers ThAr, FPI and LFC spectra.  The design of the HARPS FPI is described in \citet{2010SPIE.7735E..4XW}. Briefly, the HARPS FPI has a cavity width of about $7.3$\,mm and a measured finesse of about 4.3. It is fiber coupled and temperature and pressure stabilized inside a vacuum tank. 

We analyzed calibration data taken on March 16, 2011, offering both ThAr and FPI exposures. We reduced the calibration spectra using the IDL REDUCE package. The pixel position of the Thorium lines were measured by fitting a Gaussian profile to the extracted 1D spectra. The FPI peaks are separated by 8 -- 21 pixels (from blue to red orders). The median FWHM of the FPI lines is about 5 pixels or 1.6  resolution elements which means the FPI lines are moderately resolved in the spectra. Hence the line shape of the FPI deviates from the Gaussian like instrumental profile of HARPS and we fit a Lorentz function to each peak to obtain the pixel position. 

\begin{figure} 
\includegraphics [width=\linewidth]{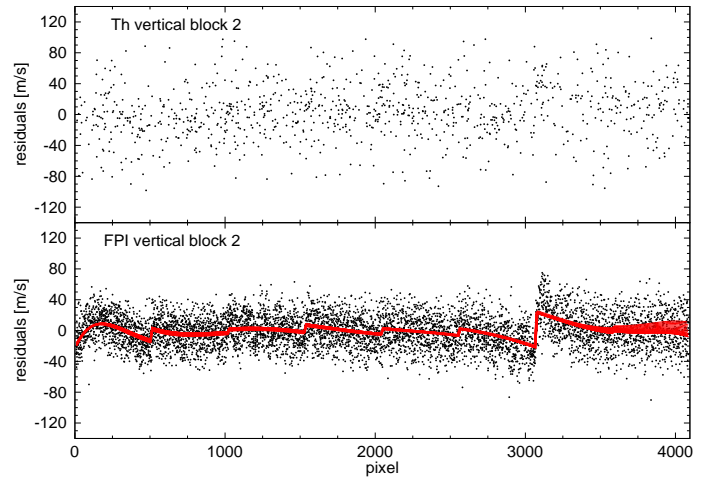} 
\caption[D_CARMENES]{Residuals of Th and FPI positions with respect to a simple polynomial wavelength solution for HARPS orders 116 – 135 (black dots). The difference of the wavelength solutions between the 
polynomial model and the model with variable pixel sizes (red lines) shows the 
captured systematics.} 
\label{fig:new_wls} 
\end{figure}

The first step in our calibration procedure is to derive the wavelength solution from the Thorium lamp only using the line list of \citet{1983ats..book.....P}. Because of the low number of Thorium lines available in single calibration images we use a $2$D polynomial in the direct regression ($7$th degree in dispersion direction and $6$th degree in cross dispersion) and uniform pixel sizes. The residuals of the best fit to the Thorium lines in block $2$ of the blue detector (echelle orders $116$ to $135$) are plotted in the upper panel of Fig.~\ref{fig:new_wls}. 
 
\begin{figure} 
\includegraphics [width=\linewidth]{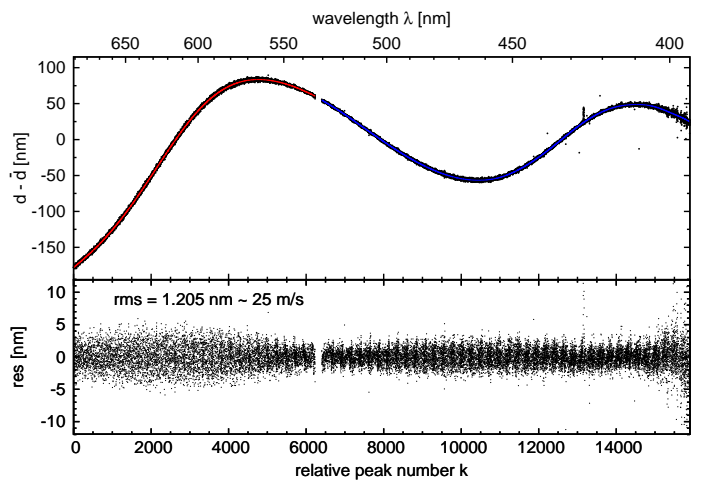} 
\caption[D_CARMENES]{Effective cavity width measurement of the HARPS FPI using the Thorium wavelength solution. Upper panel: Measured effective cavity width (black dots), spline fit to red and blue HARPS CCDs (red and blue line, respectively). Lower panel: Residuals of the spline fit. Orders $89 - 153$ are shown here.} 
\label{fig:HARPS_D} 
\end{figure}

This HCL wavelength solution was used as input for the FPI calibration to assign wavelengths to all FPI peaks. We chose $m_1=27526$ ($\lambda_1 = 5304.24$\,$\AA$) which resulted in the minimum penetration depth variation. The model for $d(k)$ was a B-spline with $23$ nodes in the blue and $13$ nodes in the red HARPS regime. The effective cavity width measurement $d_k = 2\, \lambda_k \,(m_1+k)$ and the according model $d(k)$ are shown in Fig.~\ref{fig:HARPS_D}. 

The mean effective cavity width we measure for the HARPS FPI is $\bar{d} = 7.300\,20 \pm 0.000\,15$~mm. (The spike seen around $430$~nm is caused by a CCD defect in the blue detector which systematically influences the line position measurement in this region.) 

From the B-spline model and the peak number we derived the calibrated FPI wavelengths. To check the quality of the Thorium wavelength solution we plot the residuals of the calibrated FPI lines included in block $2$ of the blue detector (echelle orders $116$ to $135$) in the lower panel of Fig.~\ref{fig:new_wls}. Due to the numerous more lines offered by the FPI we are now able to see systematics that were hidden in the Th residuals simply due to the lack of lines. 

These systematics are already known from laser frequency comb calibration and are due to a stitching effect \citep{2010MNRAS.405L..16W}.
The $4$k~$\times$~$4$k HARPS detector consists of two $4$k~$\times$~$2$k CCDs; each is build up of $8$~$\times$~$2$ blocks with $512 \times 1024$ pixels.

 To capture the effect caused by the CCD imperfections in our
wavelength solution, we applied our pixel size concept (see Sect.~3.2)
and model the size of every $512$th pixel as indicated in Eq.~(\ref{eq:pixsizes}). From the HARPS flat fields we see that the first pixel block is 3 pixels smaller than the rest and ends at pixel $509$ while the rest continues in $512$ pixel intervals. The edge of the detector is a small block of $3$ pixels which is not fitted. Furthermore, we assume
that the last pixels in the same column have the same size which allows us to
limit the number of free pixel parameters to 14 per CCD. We found it
necessary to increase the polynomial orders ($8$th degree in dispersion
direction and $7$th degree in cross dispersion). This is possible due to
the numerous FPI lines.

We overplot the model difference between the simple polynomial and the full model including inhomogeneous column sizes as solid red lines in the lower panel of Fig.~\ref{fig:new_wls}. Our new wavelength solution handles most of the systematics a simple polynomial model cannot account for. Hence using the FPI along with a wavelength solution including unequal pixel or column sizes is a powerful method to characterize the detector and calibrate the spectrograph at the same time. 

Table~\ref{tab:pixel_sizes} shows the deviation of column sizes from unity, $a_{\mathrm{pix}} - 1$. The smallest column we found differs by $5.8\%$ is size which translates to about $46$~m/s for HARPS.

\begin{table*} 
      \caption[]{Deviation of column sizes, $a_{\mathrm{pix},i}$ from unity in percent for HARPS. Estimated errors are typically $0.3\%$ of a pixel ($\sim\,2.5$~m/s).} 
         \label{tab:pixel_sizes} 
\setlength{\tabcolsep}{17.pt} 
\begin{tabular}{cccccccc} 
\hline  \hline 
\begin{tabular}[c]{@{}c@{}}Column number $i$\end{tabular}             & 509   & 1021  & 1533  & 2045  & \multicolumn{1}{l}{2557} & \multicolumn{1}{l}{3069} & \multicolumn{1}{l}{3581} \\ \hline 
\begin{tabular}[c]{@{}c@{}}Block row 1 ($o = 135-161$)\end{tabular} & -4.3 & 1.1 & -1.1 & -3.5 & 0.5 & -4.5   & 1.3                    \\ 
\begin{tabular}[c]{@{}c@{}}Block row 2 ($o = 116-135$)\end{tabular} & -2.2 & -0.8 & -1.4 & -1.0 & -1.5  & -5.8  & -0.5                    \\ 
\begin{tabular}[c]{@{}c@{}}Block row 3 ($o = 100-114$)\end{tabular} & -5.3 & 2.4 & -3.1 & -3.4 & -0.5  & -4.6  & 1.6                    \\ 
\begin{tabular}[c]{@{}c@{}}Block row 4 ($o = 89-99$)\end{tabular}   & -2.8 & -1.0 & -5.0 & -0.4 & 0.2   & -5.8 & -0.2            \\ \hline       
\end{tabular} 
\end{table*} 

\subsection{Comparison to DRS} 

To check the accuracy of our calibration using the HARPS FPI we compare our wavelength solution to the standard calibration delivered by HARPS DRS (Data Reduction Software). The difference between our wavelength solution and the cubic polynomial reverse regression of DRS is shown in the upper panel of  Fig.~\ref{fig:LFC_vs_FPI}. We observe distortions with amplitudes of up to $50$\,m/s that are 
repeating in every order. This is very similar to the results found in \citet{2010MNRAS.405L..16W} and \citet{2013A&A...560A..61M} using LFC data. These authors argue that the 
pattern originates from DRS not taking into account different sizes of 
detector pixels. The difference between the LFC solution of \citet{2013A&A...560A..61M} and the DRS solution yields an rms scatter of $25$\,m/s. We confirm the distortions in the DRS solution with the use of the HARPS FPI and find an rms scatter between our wavelength solution and the DRS solution of $22$\,m/s.

\subsection{Comparison to LFC}
 
\begin{figure} 
\includegraphics [width=\linewidth]{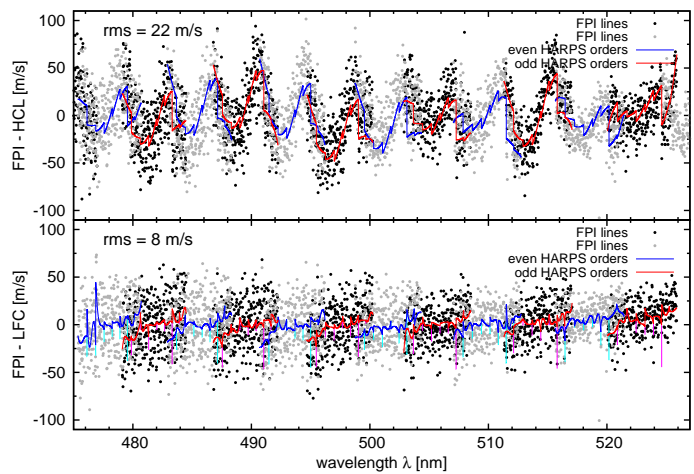} 
\caption[D_CARMENES]{Upper panel: Difference between our FPI wavelength solution and the DRS wavelength solution (solid lines) with an rms of $22$\,m/s. Difference between FPI wavelengths obtained from the model $d(k)$ and the wavelengths assigned to the FPI from DRS wavelength solution (points). Lower panel: Same as above but using the LFC wavelength solution of \citet{2013A&A...560A..61M}. Orders $116 - 128$ are shown here. } 
\label{fig:LFC_vs_FPI} 
\end{figure}

It is also interesting to compare our wavelength solution directly to the LFC solution. Unfortunately, there is no night in the HARPS archive with FPI and LFC spectra. Hence we compute the difference between the LFC wavelength solution of \citet{2013A&A...560A..61M} for November 24, 2010 and our FPI wavelength solution for March 16, 2011 and correct for $6$~m/s which is the mean drift between both Thorium solutions. The result is plotted in the lower panel of Fig.~\ref{fig:LFC_vs_FPI}.

The deviations between our FPI wavelength solution and the LFC solution 
of \citet{2013A&A...560A..61M} are smaller, only $8$\,m/s. Generally, no large amplitude distortions of 
$50$\,m/s are visible as it is the case with the ThAr solution.

We comment now on some remaining features. The peak around $477$\,nm is due 
to the low flux level of the LFC at this wavelength.

In six pixel wide regions at the block borders (marked in cyan and 
magenta in the lower panel of Fig.~\ref{fig:LFC_vs_FPI}) the deviations can jump by 50 m/s.
This is due to an indexing problem. We realized that the 
discontinuities (with 512 pixel period) start at pixel 515 in the LFC 
solution of \citet{2013A&A...560A..61M}. We chose 509 as start pixel to 
match the columns having anomalous sensitivity in the flat echelle image 
likely due to their pixel size.

Apart from this deviations we observe a residual trend 
across the orders which is not well understood, but we note that four months 
passed between both solutions.
Around this trend differences with amplitudes of about 10 m/s are seen. 

With the 
current data quality of the HARPS FPI spectra we cannot completely 
resolve all the fine structure that is seen with the LFC. \citet{2013A&A...560A..61M} 
fit a cubic polynomial for each block in each order individually, 
i.e. their model has more parameters and is capable of handling more 
small scale structure. We used a less flexible model due to the problems with the HARPS FPI spectra 
discussed in detail in Sect. 4.4. If the data quality 
improves, we might also be able to use more detailed models and to 
characterize the detector in more detail.

%
\subsection{Systematic high frequency noise in the HARPS etalon} 

When we calibrated the HARPS FPI the rms scatter around the B-spline model of $d(k)$ was $25$~m/s, see Fig.~\ref{fig:HARPS_D}. This is about one order of magnitude more than what we expect from photon noise for signal to noise ratios of 150 to 300. We found that the residuals are not white but exhibit high frequency variations. The position residuals of lines present in adjacent and overlapping orders are correlated indicating a wavelength dependency. Frequency analysis revealed two significant periods. For order $131$ we found $P_1 = 1.5056 \pm 0.0017~\AA$ and $P_2 = 0.6612 \pm 0.0005~\AA$. Furthermore, periodic peak height variations of up to $10$\% are
visible in the 1D extracted FPI spectra. Frequency analysis on the
derived FPI amplitudes yields two significant periods matching
those found in the position residuals. We presume that an additional low finess interference signals, caused by
plane parallel optical elements in the path of the FPI like filters or
by the thickness of the FPI mirrors itself, superimposes on the HARPS FPI (\citet{2010SPIE.7735E..4XW}, Pepe, private communication). The periods would correspond to FPIs with cavity widths in
the order of millimeters. Further we find that the periods decrease towards blue and increase towards red orders, indicating that the effective cavity width of the optical element causing the additional FPI effect changes with wavelength; this can be caused by a wavelength dependent refractive index $n$. We presume that the slopes due to the flux variations shift the photo-center of the FPI lines resulting in systematics in the line position measurement. 

Without exact knowledge about the cause of this effect, our attempts to fit and correct the flux variations present in the spectra gave unsatisfying results. Therefore the FPI wavelengths are not known precise enough to resolve the fine structure that is seen in the LFC. Our method involving the FPI in the wavelength calibration for HARPS is thus limited by this systematic high frequency noise. Our laboratory experiment in Sect.~2.2 is limited by photon noise ($8$~m/s). With this data quality we would be able to resolve more details in the wavelength solution systematics of HARPS. For future projects it is advisable to avoid optical elements in the light path of the FPI that might cause additional interference patterns and to characterize the FPI in advance with other instruments. 

%

\section{Influence of distortions in the wavelength solution on precise RV measurements} 

To estimate the influence of distortions in the wavelength solution on precise RV measurements with high resolution echelle spectrographs we simulate observations taken over the period of one year. The simulated spectrograph has 70 spectral orders each covering $57\,\AA$, a delta peak instrumental profile and a detector with 4096 homogeneous pixels. The observed star is perfectly quiet and has a zero RV signal. Its spectrum consists of randomly distributed delta peaks (100 lines per order). Furthermore, the star lies in the ecliptic resulting in a $30\,000$\,m/s RV signal from the barycentric motion of Earth. This oversimplified approach allows us to single out the effect of the wavelength solution only on our observations without additional noise from any other sources. 

\begin{figure} 
\includegraphics [width=\linewidth]{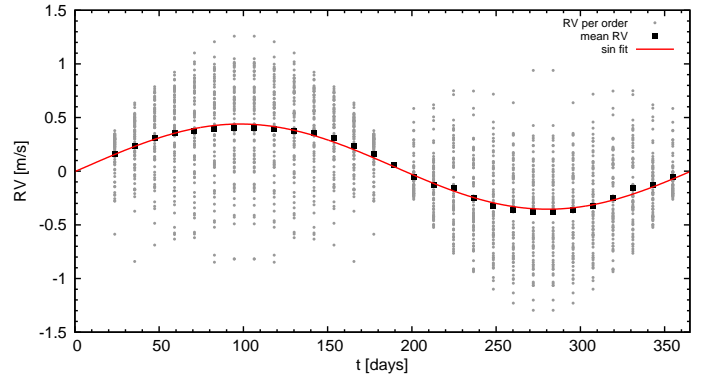} 
\caption[D_CARMENES]{Simulation of the RV signal due to a disturbed wavelength solution. RV signal for single orders (gray dots), mean RV of all orders (black squares), sinusoidal fit to the mean RV (red line)} 
\label{fig:rv_sim} 
\end{figure} 

In total we simulate $29$ observations over one year. We initialize the stellar spectrum by computing random wavelengths such that 100 delta peaks fall on each of the 70 orders of the spectrograph. At time stamp $0$ the barycentric motion of Earth towards the star is zero; hence this spectrum represents the rest frame. The spectra for all observation time stamps are computed applying only the Doppler shift corresponding to the barycentric motion of Earth around the Sun ($\sim 30\,000$\,m/s). We ignore the effect of Earths rotation because it is small compared to the motion of Earth around the Sun ($\sim 500$\,m/s). Spectra are taken every 12.175 days. We use the inverted wavelength solution of the spectrograph to place the stellar lines on the detector and to obtain our simulated observations. 

If the true wavelength solution of the simulated spectrograph is used to convert the pixel value back to wavelength and the barycentric motion of Earth is corrected, we obtain RVs of $0$\,m/s for all stellar lines within machine precision. 

Next we disturb the wavelength solution of the spectrograph. We want to test the effect of small systematics in the wavelength solution on precise RV measurements. Hence, the difference between our Thorium wavelength solution and our calibrated FPI wavelength solution in HARPS order 118 (Sect.~4.1, red line in Fig.~\ref{fig:new_wls}) is added to the true wavelength solution of the simulated spectrograph. 

Now we use the disturbed wavelength solution, convert the pixel values into wavelengths, correct for the barycentric motion of Earth and compute the RVs for all lines. We average the RV of single lines to compute the RV order by order for each observation (gray dots in Fig.~\ref{fig:rv_sim}). The RV per time stamp is the mean of the individual order RVs (black squares in Fig.~\ref{fig:rv_sim}).

We observe the following effects in Fig.~\ref{fig:rv_sim}. The RVs derived for individual orders (gray dots) show a scatter of about $1$\,m/s. This effect is a result of the random distribution of lines on the CCD and the systematics in the wavelength solution.

 Depending on the location of a stellar line on the CCD it will give a different systematically wrong RV because the systematic error in the wavelength solution is different for different pixels. We can derive the expected RV signal $\delta_v$ resulting from the wavelength solution systematics $\delta_\lambda(x)$ by inserting the wavelength solution distortion in the Doppler equation: 
   \begin{equation} 
      \frac{[\lambda(x_i) + \delta_\lambda(x_i)] - [\lambda(x_0) + \delta_\lambda(x_0)]}{\lambda(x_0) + \delta_\lambda(x_0)}=\frac{v + \delta_v}{c} 
      \label{eq:doppler_dist} 
   \end{equation} 
where $x_0$ and $x_i$ correspond to the line positions on the CCD at time stamps $t_0=0$ (rest frame) and $t_i$ respectively. The barycentric motion of Earth shifts lines on scales of a few 10 pixels. On this scale the change in $\delta_\lambda$ is small and we can approximate the difference $\delta_\lambda(x_1)-\delta_\lambda(x_0)$ with the derivative of the wavelength solution distortion multiplied by the pixel shift $\mathrm{d}\delta_\lambda(x) / \mathrm{d}x \cdot \Delta x$. The pixel shift $\Delta x$ depends on the barycentric velocity of Earth $RV_E$ and the pixel scale in terms of RV $\kappa$. For HARPS one pixel covers approximately $800$~m/s/pix. Hence a barycentric velocity of $RV_E = 30\,000$~m/s shifts the lines by $37.5$ pixels on the detector. Finally the small term $\delta_\lambda(x_0)$ in the denominator can be neglected and velocity signal of a single line $\delta_v$ due to the distortion in the wavelength solution $\delta_\lambda(x)$ can be written as
   \begin{equation} 
      {\delta_v} = \frac{c}{\lambda(x_0)} \cdot \frac{d \delta_\lambda(x)}{dx} \cdot \frac{1}{\kappa} \cdot RV_E\ .
     \label{eq:derived_rv} 
   \end{equation} 
Averaging over all the different $\delta_v$ in one order results in a systematic RV that depends on the random distribution of stellar lines on the CCD. As the distortion is the same for all orders, the simulation results presented in Fig.~\ref{fig:rv_sim} are equivalent to repeating the test 70 times for one order with different random sets of lines. Hence we observe scatter between the RVs of different orders only caused by the line distribution on the detector. This scatter is mainly caused by lines experiencing different slopes in ${\mathrm{d} \delta_\lambda(x)}/{\mathrm{d}x}$ rather than by single lines moving across the jumps in the wavelength solution.

The slope ${\mathrm{d} \delta_\lambda(x)}/{\mathrm{d}x}$ is, however negative almost everywhere on the CCD. Hence many lines experience a systematic RV shifts in the same direction. Therefore, the mean RV derived from the individual orders shows a low amplitude RV signal with a period of one year. The amplitude of this signal is about $0.5$~m/s. Advancing to cm/s precision we cannot afford systematic errors in the order of $0.5$~m/s to $1$~m/s. Hence, accurate wavelength calibration is critical for reaching the aims with the next generation of instruments. 

%

\section{Summary and conclusions} 

HCLs have been calibration sources for echelle spectrographs since decades but advancing to cm/s RV precision introduces the need for more suitable standards. Aside from the LFC, FPIs also offer a dense comb pattern of lines that can be used for calibration.

 Because the cavity width is not known accurately enough and possibly 
drifting over time, the peak wavelengths are uncertain and FPIs cannot 
be used as a standalone calibrator. Therefore we developed a method to 
anchor the dense etalon lines with the absolute HCLs to combine their 
precision and accuracy.
Central to our method is that we model the cavity width (including its 
global wavelength dependency) in the calibration procedure. We exploit 
the assumption that the wavelength dependency of the cavity width is 
smooth which enhances the quality of the wavelength solution on medium 
to small scales, while the global accuracy is still determined by the 
HCL. The assumption of smooth cavity width was verified with FTS 
measurements on CARMENES FPI.

Along with our model for the FPI, we described our model for the 
wavelength solution of the spectrograph. Our four main points for the 
calibration are:

\begin{itemize} 
\item We suggest to perform direct regression with $x(\lambda)$ rather than reverse regression $\lambda(x)$ because the independent quantity is the wavelength $\lambda$ and not the pixel position $x$ of the spectral line on the detector.
\item We use a 2D polynomial as model in the direct regression. This couples orders, reduces the number of parameters and enhances the robustness of the fit. 
\item Motivated by the grating equation we parametrize our model as $x(\lambda,o)=\mathrm{poly}(o\lambda,o)$ which is a better representation of the problem than $\mathrm{poly}(\lambda,o)$. 
\item We account for inhomogeneous detector pixel sizes $a_i$ and transform the observed pixel positions $X$ to focal plane coordinates $x(X,a_i)$ before performing direct regression. 

\end{itemize} 

Using this concept for the HARPS FPI we are able to characterize the detector by taking into account pixel stitching and compute the wavelength solution simultaneously. 

The standard DRS HCL calibration deviates from the LFC solution with an rms scatter of $25$\,m/s. Direct comparison of our FPI wavelength solution with the LFC yields only small differences of rms $= 8$\,m/s. Hence we show that the calibrated FPI is a suitable calibration source for high resolution echelle spectrographs. 

Finally, we performed simulations to derive the fundamental limits an imperfect wavelength solution imposes on RV measurements and find that even small distortions in the wavelength solution lead to systematics signals and additional noise, both in the order of $1$ m/s. We conclude that wavelength calibrators, such as LFCs or FPIs, are necessary that provide a much denser line grid than HCLs do. We suggest that a calibrated FPI can be considered as economical alternative to LFCs to achieve very high RV precision for current and future instruments. 
 
\begin{acknowledgements} 
We thank the unknown referee for careful reading and helpful 
comments.
We thank Sebastian Sch\"afer, Ulrike Lemke and Spencer Carmichael for their help in acquiring FTS data of the CARMENES NIR FPI. Florian Franziskus Bauer acknowledges support from the Deutsche Forschungsgemeinschaft under DFG GrK 1351. 
     Mathias Zechmeister acknowledges support by the European Research Council under the FP7 starting Grand agreement number 279347. Ansgar Reiners acknowledges funding through a Heisenberg professorship under DFG 
RE-1664/9-2. 
\end{acknowledgements}


\bibliographystyle{aa}

\end{document}